\documentclass[10pt, doublecolumn]{IEEEtran}
\usepackage{graphicx}
\usepackage{lipsum}
\usepackage{amsmath,amssymb}
\usepackage[font=scriptsize]{caption}
\usepackage{citesort}
\usepackage{fancyhdr}
\usepackage{mdwmath}
\usepackage{mdwtab}
\usepackage{balance}
\usepackage{xcolor}
\usepackage{bm}
\usepackage{float}
\usepackage{amsthm}
\usepackage{multirow}
\usepackage{flafter}
\usepackage{caption}
\usepackage{subcaption}
\usepackage{setspace}
\usepackage{cite}
\usepackage{algpseudocode}
\usepackage{algorithm}
\usepackage{multirow}
\usepackage{flafter}

\DeclareMathOperator*{\argmax}{argmax}

\newtheorem{remark}{Remark}

\newtheorem{theorem}{Theorem}

\newtheorem{lemma}{Lemma}

\newtheorem{corollary}{Corollary}

\newtheorem{assumption}{Assumption}

\theoremstyle{definition}

\begin{document}

\title{A Power-Pool-Based Power Control in Semi-Grant-Free NOMA Transmission}
\author{Muhammad~Fayaz,~\IEEEmembership{Graduate Student Member,~IEEE,}
       Wenqiang~Yi,~\IEEEmembership{Member,~IEEE,}
        Yuanwei~Liu,~\IEEEmembership{Senior Member,~IEEE,}
        and Arumugam~Nallanathan,~\IEEEmembership{Fellow,~IEEE}
\thanks{M. Fayaz, W. Yi, Y. Liu, and A. Nallanathan are with Queen Mary University of London, London, UK (email:\{m.fayaz, w.yi, yuanwei.liu, a.nallanathan\}@qmul.ac.uk).\newline
	\indent M. Fayaz is also with the Department of Computer Science and  IT, University of Malakand, Pakistan.}
}

\markboth{This work has been submitted to the IEEE for possible publication. Copyright may be transferred without notice.}
{}

\maketitle

\begin{abstract}
In this paper, we generate a transmit power pool (PP) for Internet of things (IoT) networks with semi-grant-free non-orthogonal multiple access (SGF-NOMA) via multi-agent deep reinforcement learning (MA-DRL) to enable open loop power control (PC). The PP is mapped with each resource block (RB) to achieve distributed power control (DPC). We first formulate the resource allocation problem as stochastic Markov game, and then solve it using two MA-DRL algorithms, namely double deep Q network (DDQN) and Dueling DDQN. Each GF user as an agent tries to find out the optimal transmit power level and RB to form the desired PP. With the aid of dueling processes, the learning process can be enhanced by evaluating the valuable state without considering the effect of each action at each state. Therefore, DDQN is designed for communication scenarios with a small-size action-state space, while Dueling DDQN is for a large-size case.  Moreover, to decrease the training time, we reduce the action space by eliminating invalid actions. To control the interference and guarantee the quality-of-service requirements of grant-based users, we determine the optimal number of GF users for each sub-channel. We show that the PC approach has a strong impact on data rates of both grant-based and GF users. We demonstrate that the proposed algorithm is computationally scalable to large-scale IoT networks and produce minimal signalling overhead. Our results show that the proposed MA-Dueling DDQN based SGF-NOMA with DPC outperforms the existing SGF-NOMA system and networks with pure GF protocols with 17.5\% and 22.2\% gain in terms of the system throughput, respectively. Finally, we show that our proposed algorithm outperforms the conventional open loop PC mechanism.
\end{abstract}
\begin{IEEEkeywords}
	Distributed power control, Internet of things, multi-agent reinforcement learning, non-orthogonal multiple access, semi-grant-free transmission
\end{IEEEkeywords}
\section{Introduction}
Non-orthogonal multiple access (NOMA) is a promising multiple access paradigm for one of the most important use cases in fifth-generation (5G) and beyond cellular networks, namely massive machine type communication (mMTC) \cite{6824752}. More specifically, providing massive connectivity for satisfying the explosive increase in mobile devices is the main challenge for mMTC. To this end, power domain NOMA becomes a suitable solution as it allows multiple users or devices to share the limited bandwidth resource rather than to solely occupying them \cite{8114722}. In particular, NOMA multiplex different users in the same time/frequency resource block (RB) by using the superposition coding at the transmitter side and the successive interference cancellation (SIC) method at receivers \cite{8353359}.
To enable the accomplishment of mMTC, as well as ensuring the quality of service (QoS) with low latency communication and small signalling overhead, NOMA with two types of access methods, i.e., grant-free (GF) and grant-based (GB) access are proposed \cite{fayaz2021transmit}. In GB transmission, a user processes some handshakes before actual data transmission which arises
signal overhead with high access latency. Besides, GB transmission is not suitable for some Internet of things (IoT) scenarios, where the IoT applications require low data rate but massive connectivity and high energy efficiency \cite{yi2020multiple}.
In GF transmission, the user directly transmits data without any prior handshakes or schedule requests \cite{8663999}. Therefore, GF transmission provides massive connectivity for short-packet IoT applications. However, GF transmission leads to frequent collisions due to the absence of base station (BS) involvement in the scheduling of orthogonal resources \cite{8454392}. Recently, a hybrid version of GF and GB NOMA known as semi-grant-free (SGF) NOMA is considered for uplink transmission due to its potential to enhance massive connectivity and reduce access latency via allowing GF and GB users to share the same RBs \cite{9244136}. It is worth noting that SGF-NOMA also ensure the QoS of GB users as it only allocates redundant resource of GB users to GF users.

To enable an open-loop with distributed power control (DPC) for SGF-NOMA IoT networks, this paper aims to exploit the potential of multi-agent deep reinforcement learning (MA-DRL) to design a transmit power pool (PP) \cite{fayaz2021transmit} associated with each RB. Each GF user receives the QoS and other information from the base station (BS) and adapts its transmit power accordingly. Furthermore, a GB user access the channel via GB protocol and GF users are allowed to access it opportunistically.

\subsection{Existing SGF-NOMA Works and Motivation}
Recently, pure GF and GB NOMA transmission schemes have been studied extensively from various aspects. However, power domain SGF-NOMA assisted methods are still in their infancy. 
In \cite{fayaz2021transmit}, the cell area is divided into different layers, and a layer-based transmit PP is designed for GF IoT networks using a cooperative MA Double Deep Q Network (DDQN) algorithm.
To ensure the QoS of GB users and to restrict the number of GF users, two techniques are proposed in\cite{8662677}. In the proposed scheme, contention control mechanisms are developed that  effectively control the number of GF users to suppress the interference to GB users. In first technique, the BS decodes the GB users signal in the first stage of SIC, i.e., only GF users with the weak channel are allowed to share the resources with GB users. In the second scheme, the BS decodes the GF users signals first, i.e., users with strong channel gain can share the resources with GB users. Therefore, the first scheme is ideal when GF users are cell-edge users. However, the second scheme is more suitable for the scenario when GF users are close to BS. In \cite{ding2020new}, the QoS of GB users has been ensured by exploiting the flexibility in choosing the NOMA decoding order. This new scheme is a combination of the two schemes discussed in \cite{8662677} with some additional advantages. As compared to the previous two schemes, the work in \cite{ding2020new} efficiently avoids error floors in outage probability and significantly increases transmission robustness without precise PC between users. The authors in \cite{zhang2020semi} investigate SGF-NOMA considering two different scenarios. In the first scenario, GF users are considered as cell-centre users and GB are considered as cell-edge users. In the second scenario, GF users are located at the edge of the cell, while GB users are distributed near the BS. To decide whether GF users are able to share the channels occupied by GB users, the authors proposed a dynamic protocol to determine the dynamic channel quality threshold to decrease the unexpected interference to GB users. The proposed dynamic protocol shows superior results for both scenarios as compared to the open-loop protocol. In \cite{9244136}, the received power of GB users is used as a dynamic quality threshold and the closed-form expressions for GF and GB users are derived. The maximum rate of GF users and the ergodic rate without error floor for GB users has been observed. An adaptive power allocation strategy for GB users is adopted in \cite{9119121} to encounter performance degradation problem caused by GF users sharing the channels with GB users. Maximum data rate based scheduling scheme is proposed in \cite{lu2020advanced}. GF users that produce more data rate are scheduled to pair with GB users for uplink transmission.

Since in modern communication systems, millions of users with heterogeneous QoS and performance requirements are expected in a single cell to deliver diverse applications. In order to satisfy the heterogeneous QoS requirements and share limited communication resources appropriately, a distributed mechanism is necessary to allocate limited resources accordingly based on the QoS requirements of the IoT devices. Therefore, IoT users must be able to acquire their communication resources autonomously, as it is impractical to expect them to interact with the BS frequently, given their limited resources \cite{park2019distributed}. Furthermore, in a large-scale IoT scenario, the BS would not be able to manage the network resources effectively and promptly. In addition, considering the distributed nature of IoT systems, distributed learning and resource allocation for IoT applications are worth to be investigated \cite{8712527}.
In the existing works, to control interference, BS plays a dominant role to increase or decrease the number of GF users. This involvement of BS increases complexity at the BS side \cite{9205230}. Additionally, a fixed power control (FPC) strategy is used with closed-loop PC, resulting in extra signal overhead and compromising users' fairness. Moreover, when the channel quality (strong or weak users) or received power threshold is used for admitting GF users, the GF users close to the BS are always scheduled to transmit with GB users, which compromises user fairness. Furthermore, open-loop PC in networks helped by SGF-NOMA is difficult to achieve with conventional optimization approaches. It's worth noting that allocating resources in time-varying cellular networks is an NP-hard problem\cite{8387468}. The optimization problem in wireless communication is also combinatorial in nature, and it is mathematically intractable as the size of the network increases. By contrast, machine learning (ML) can solve NP-hard optimization problems more efficiently than traditional optimization and brute-force methods\cite{9097306}\cite{7792374}. As a substitute for equations and models, ML techniques observe hidden patterns in data to make near-optimal and best possible decisions. It is desirable to use ML-based algorithms in wireless communications and beyond, especially for mMTC, because the complexity of such processes increases exponentially with the number of users in the network \cite{9381701}. In particular, reinforcement learning (RL) has the potential of taking decisions and performing learning simultaneously, which is one of the perfect features for wireless communication applications \cite{luong2019applications}. Therefore, we adopt an ML-based method to solve the considered problem since it can provide excellent approximate solutions while dealing with extensive state and action spaces.
\subsection{Contributions}
Motivated by the aforementioned issues, this paper focuses on the design of PP for each RB to enable an open-loop with DPC. To deal with the intra-RB interference problem and meet the QoS requirements of GB users, we utilize MA-DRL to find the optimal transmit power levels for each PP to enable open-loop PC. The main contributions are outlined as follows:
\subsubsection{Novel MA-DRL Framework for SGF-NOMA} We propose a novel MA-DRL based SGF-NOMA algorithm, which is the first scheme in SGF-NOMA IoT networks in the literature that uses DRL for power and sub-channel selection. Moreover, we formulate the throughput optimization problem as a multi-agent Markov decision process (MA-MDP) and solve it using a MA-DDQN and MA-Dueling DDQN algorithm instead of a conventional DQN algorithm to overcome the problem of overestimating Q values and to enhance agents learning efficiency.
\subsubsection{Autonomous and Distributed PC via Power Pool} We design a transmit power pool for each sub-channel and assign it to the corresponding sub-channel, which enables an open-loop with a DPC strategy.
To obtain optimal transmit power levels for each power pool, GF users in the network act as agents and learn a policy that guides them to adaptively adjust their transmit power and select a sub-channel (under some practical constraints) to maximize network throughput and secure GB users' QoS requirements.
Additionally, the proposed algorithm is executed without message exchange or online coordination among the users (agents).
\subsubsection{Computational Scalability and Signal Overhead} We show that our proposed algorithm is more scalable and flexible in terms of the complexity and signal overhead. Moreover, to reduce the training time, we eliminate the transmit power level (invalid actions) that cannot be opted for uplink transmission due to users transmit power constraint. Therefore, users with limited action space easily find out the best actions.
\subsubsection{Performance Analysis}
The numerical results show that MA-DDQN performs similarly to MA-Dueling DDQN in networks with a small set of states and actions. However, MA-Dueling DDQN is more efficient than MA-DDQN in networks with large action and state spaces. Furthermore, our proposed algorithm outperforms the existing SGF-NOMA systems and pure GF-NOMA IoT networks in terms of throughput. In addition, we show that the number of GF users has a positive correlation with the cluster and system throughput and find out the optimal number of GF users for each cluster. Finally, we show that our proposed algorithm outperform the conventional open loop PC mechanism in SGF-NOMA systems.

\subsection{Organization}
The rest of this paper is organized as follows. Section II outlined the system model. The DPC mechanism and problem formulation are introduces in Section III. the MA-DRL framework for SGF-NOMA systems is presented in Section IV. Section V shows numerical results. Finally, the conclusion is given in Section VI.

\section{System Model}
\begin{figure*}[t!]
	\centering
	\includegraphics[width = 1 \linewidth,keepaspectratio]{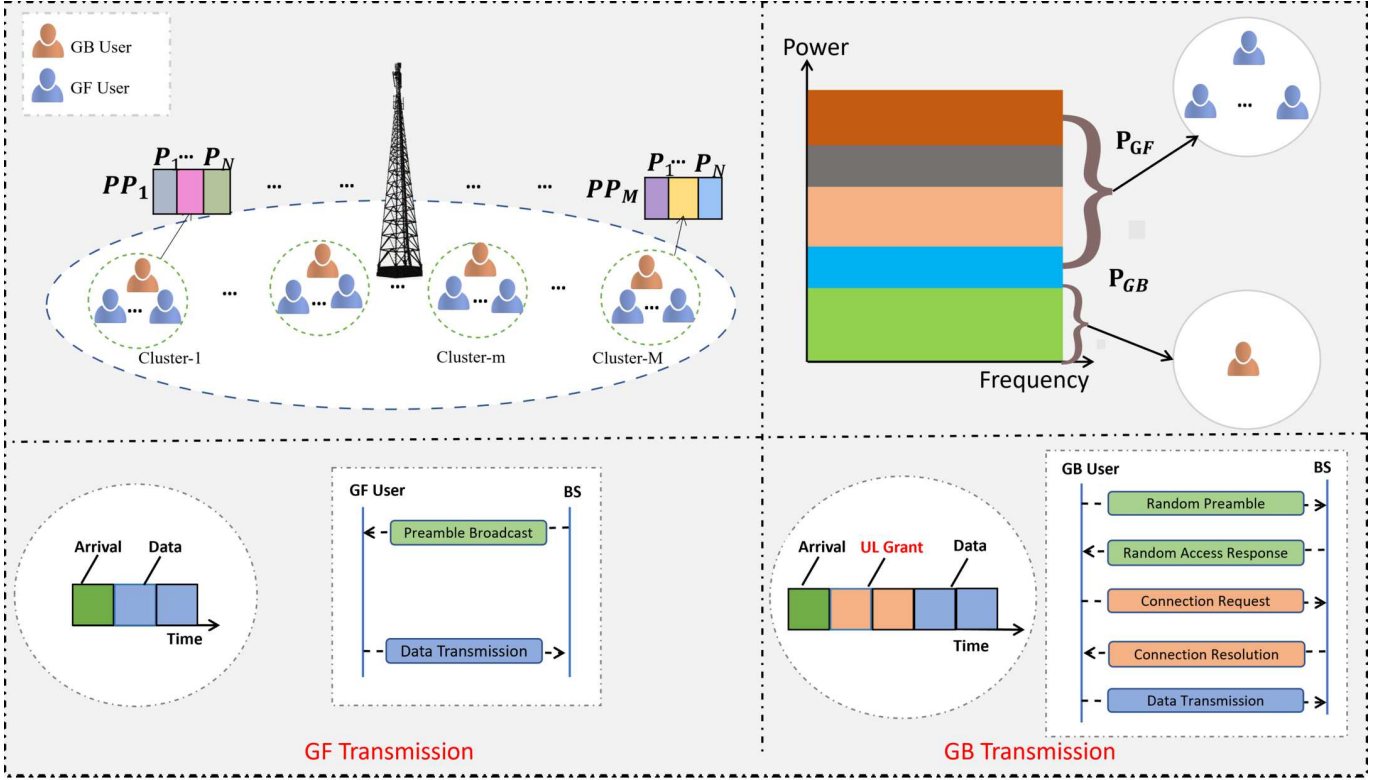}\\
	\caption{An illustrative structure of SGF-NOMA IoT networks: Top-left sub-figure shows the distribution of GF and GB users in different clusters with PPs. Top-right sub-figure represents the transmit power of GF and GB users on the same sub-channel. The bottom-left sub-figure shows the GF transmission, and bottom-right sub-figure shows the GB procedure.}
	\label{system_model}
\end{figure*}

We consider SGF transmission in IoT networks as shown in Fig.~\ref{system_model}, where a single BS is located at the geographic centre with a radius $R$. We assume that two types of users equipped with a single antenna are randomly distributed, i.e., GB and GF users that transmit uplink data to the BS. The set of users
$\mathbf{U}=\{1,2,...{N_{GF}}\}$ represents GF users, whereas GB users are denoted by $\mathbf{V}=\{1,2,...{N_{GB}}\}$. The locations of GF and GB users are modelled as two homogeneous Poisson
point processes (HPPPs) with densities $\lambda_{GF}$ and $\lambda_{GB}$. Therefore, the number of GB and GF users follows a Poisson distribution. At one time slot $t$, the probability of the number of active users $N_G$ (where, $G \in\{GB, GF\}$) equalling to $N_t\ge 0 $ is given by
\begin{align}
\Pr\{N_G=N_t\} = \frac{\lambda_G^{N_t}\exp(-N_t)}{N_t!}.
\end{align}
The probability density function of a random user with distance $r_G$ is given by $f_r (r_G) = \frac{2r_G}{R^2}.$
We define the channel gain and transmit power of $i$-th GB users as $h^{GB}_i= |h_i|^2 (r_{i,{GB}})^{-\alpha}$ and $P^{GB}_i$ respectively. Similarly, the channel gain of $j$-th GF user with transmit power $P^{GF}_j$ is given as $h^{GF}_j= |h_j|^2 (r_{j,{GF}})^{-\alpha}$. The $h_i$, $h_j$, $r_i$, $r_j$, and $\alpha$ are the small-scale Rayleigh fading of user $i \in\mathbf{V}$ and user $ j \in \mathbf{U}$, communication distances of user $i$ and user $j$, and path loss exponent, respectively. 

\subsection{SGF-NOMA Transmission}
Note that a large portion of IoT users in mMTC do not require ultra-high data rates \cite{yi2020multiple}. The conventional GB transmission based on prior handshakes with the BS, provides limited connectivity and more capacity for most IoT applications than required. This extra capacity can be utilized to enhance the connectivity via GF transmission that forms a hybrid version known as SGF-NOMA transmission. More specifically, in SGF-NOMA scheme, GB and GF users share the same or a part of the same RB for uplink transmission. Assuming the total number of orthogonal RBs (sub-channels) is $M$, the combined information received at the BS in a time slot $t_s$ is
\begin{multline}
y (t_s) = \sum_{m=1}^{M} \sum_{i=1}^{N_{GB,m}}\sqrt{P^{GB}_{m,i}(t_s)}h^{GB}_{m,i}(t_s) x_{m,i}(t_s) + \\ \sum_{m=1}^{M} \sum_{j=1}^{N_{GF,m}}\sqrt{P^{GF}_{m,j}(t_s)}h^{GF}_{m,j}(t_s) x_{m,j}(t_s) + n_{0}(t_s),
\end{multline}
where $N_{GB,m}$ and $N_{GF,m}$ are the numbers of GB and GF users in the $m$-th sub-channel, respectively. In the $m$-th sub-channel, the $x_{m,i}$ is the transmitted signal from the $i$-th GB user  and the $x_{m,j}$ is that from the $j$-th GF user. The $n_{0}$ is the additive white Gaussian noise for each sub-channel with variance $n_0^2$. At each time slot, we consider a static channel for each user, but the value of this parameter changes to new independent values during the following time slots.
\subsection{Signal Model}
We assume that the GB users have the highest priority (e.g., a sensor for healthcare monitoring) and provide the strongest received power at the receiver. Thus, the BS always decode the GB user in the first stage of SIC\footnote{In this work, we consider perfect SIC. In this case, NOMA achieves higher spectral efficiency.  However, there are several issues in practical implementation, such as error propagation; therefore, perfect SIC assumption is unfeasible in real scenarios. This challenge is out of the scope of this paper.} to avoid long latency. After that, the BS turns to decode the GF users according to the received power strength order~\cite{8114722}. To simplify the analysis, this work considers a typical scenario that each RB has one GB user\footnote{We can multiplex more than one GB user in each NOMA cluster; however, if we group more than one GB user into one NOMA cluster, we are required to satisfy the QoS of multiple GB users.}. Therefore, the received power strength order, i.e., the decoding order, at the BS can be expressed as 
\begin{multline}\label{received_power}
P^{GB}_{m,1}h^{GB}_{m,1}(t_s) \ge P^{GF}_{m,1}h^{GF}_{m,1}(t_s) \ge P^{GF}_{m,2}h^{GF}_{m,2}(t_s) \cdots \ge\\ P^{GF}_{m,N_{GF,m}}h^{GF}_{m,N_{GF,m}}(t_s). 
\end{multline}

The signal-to-interference-plus-noise ratio (SINR) for the $i$-th GB user on sub-channel $m$ in time slot $t_s$ is given by
\begin{align}
\gamma^{GB}_{m,i}(t_s) = \frac{P^{GB}_{m,i}h^{GB}_{m,i}(t_s)}{\sum_{j=1}^{N_{GF,m}}P^{GF}_{m,j}h^{GF}_{m,j}(t_s)+n_0^2},
\end{align}
whereas the SINR of the $j$-th GF user can be given as
\begin{align}
\gamma^{GF}_{m,j}(t_s) = \frac{P^{GF}_{m,j}h^{GF}_{m,j}(t_s)}{\sum_{j'=j+1}^{N_{GF,m}}P^{GF}_{m,j'}h^{GF}_{m,j'}(t_s)+n_0^2}.
\end{align}

To guarantee the SIC process and maintain the QoS of GB and GF users, the following constraints are applied:
\begin{subequations}\label{target}
	\begin{align}
	R^{GB}_{m,i}{(t_s)}=B_s\log_2(1+\gamma^{GB}_{m,i}(t_s)) \ge \tau,\\
	R^{GF}_{m,j}{(t_s)}=B_s\log_2(1+\gamma^{GF}_{m,j}(t_s)) \ge \bar{\tau},
	\label{th}
	\end{align}
\end{subequations}
where $R^{GB}_{m,i}{(t_s)}$ and $R^{GF}_{m,j}{(t_s)}$ is the data rate of GB users $i$ and GF user $j$ in time slot $t_s$, respectively. Furthermore, $\tau$ is the required target data rate to ensure QoS of GB users and $\bar{\tau}$ is the target threshold for GF users. The $B_s$ is the bandwidth of each sub-channel obtained from $B_s=B/M$, where $B$ is the total bandwidth. Although (\ref{th}) is not necessary for GF transmission, it is important for the PP design since this constraint is able to limit the number of potential GF users for each RB, which enhance the connectivity of GF users.

\subsection{Cluster Based Transmit Power Pool Generation}
\begin{figure*}[t!]
	\centering
	\begin{subfigure}{.7\linewidth}
		\centering
		\includegraphics[height= 2.6in, width=4.5in]{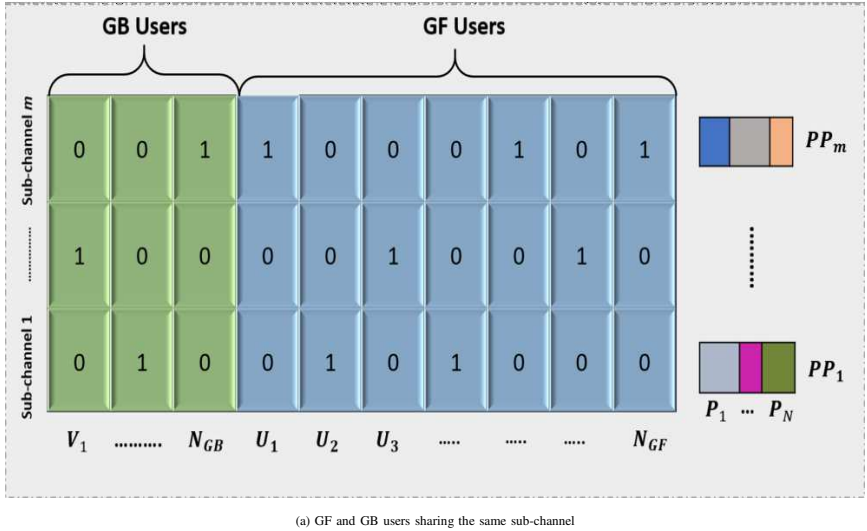}
		\caption{GF and GB users sharing the same sub-channel }
		\label{p}
	\end{subfigure}%
	\begin{subfigure}{.3\linewidth}
		\centering
		\includegraphics[height=2.55in, width= 2.0in]{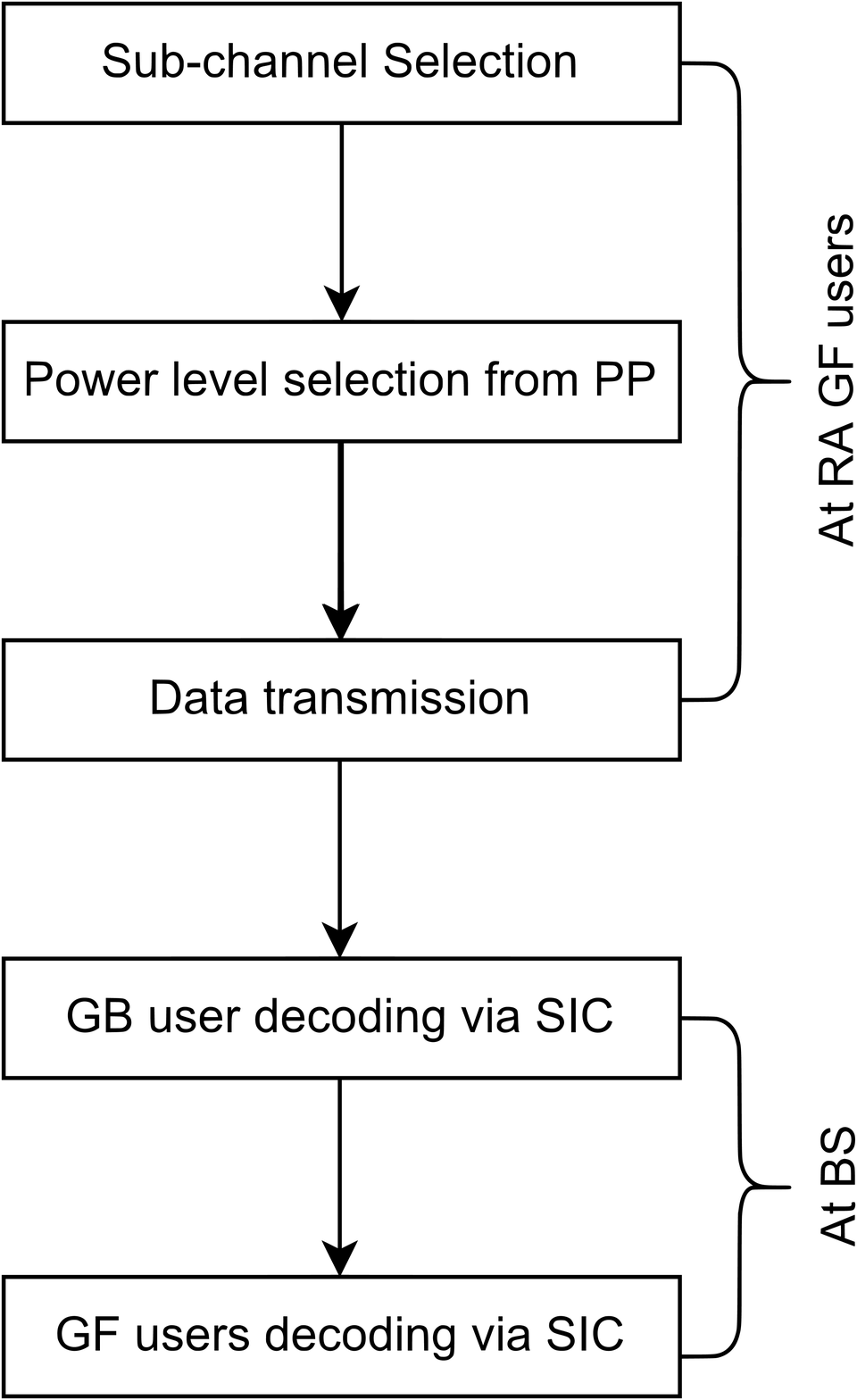}
		\caption{ PP and power level selection process}
		\label{us}
	\end{subfigure}%
	\caption{An illustrative structure of GF and GB users sharing the same sub-channel and PP selection process. In a time slot $t_s$, a user can only select one sub-channel and randomly select one transmit power level from the corresponding $\mathbf{ PP }$: Sub-figure (a) represents  GF and GB users sharing the same sub-channel and $\mathbf{ PP }$ associated to each sub-channel. Sub-figure (b) shows the PP and power level selection process at user side and SIC process at the BS side.}
	\label{PP}
\end{figure*}

Users in conventional SGF-NOMA transmit with the fixed power, and finding the optimal power for each user requires closed-loop PC, which is costly for IoT applications. This paper demonstrates the capability of MA-DRL techniques to achieve open-loop PC with DPC by generating a PP for each sub-channel, i.e., $\{\mathbf{PP}_1, \mathbf{PP}_2, \dots, \mathbf{PP}_M\}$, where, $\mathbf{PP}_m \subset \mathbf P_t = \{P_1, P_2, \cdots, P_{NP}\}$ as shown in Fig. \ref{PP}(\subref{p}). To generate PP for a sub-channel, the MA-DRL algorithm learns a policy that guides all IoT users to adjust their transmit power under some practical constraints. We assume that the BS broadcasts these $\mathbf{PPs}$ to GF users in the network. GF users then decides the sub-channel and randomly  choose the transmit power level from the $\mathbf{PP}$ corresponds to the selected sub-channel for uplink transmission. For example, if a GF user $j$ wants to transmit on sub-channel 1,  it selects a transmit power level from $\mathbf{ PP}_1$ for uplink transmission. Selecting a transmit power level from $\mathbf{ PP }$ corresponds to each sub-channel restrict the interference to a tolerable threshold $\phi$ that ensures the QoS of the GB users. Fig. \ref{PP}(\subref{us}) shows the PP and power level selection process at user side and SIC process at the BS side.

\begin{remark}
	These PPs can enable GF users to transmit with optimal power level without any training. More specifically, new users joining the network receive these PPs (optimal power levels) from the BS and they can select power for transmission without any prior training which prevent training complexity and reduce energy consumption.
\end{remark}
\section{Autonomous Power Control and Problem Formulation}
In SGF-NOMA, the GF users are allowed to access channels occupied by GB users. More specifically, we assume that a single GB user $i$ is connected to the BS through sub-channel $m$. 
Let $\mathbf{Q}^m_q$ represent the number of GF users share the same sub-channel $m$ with GB user and can be expressed as $\mathbf{Q}^m_q\subset \mathbf{U}=\{q: 0\leq q \leq N_{GF}\}$. To control GF users transmit power and to restrict interference for maintaining uplink QoS of GB users, the central BS has three steps:
\begin{itemize}
	\item The BS acquire complete CSI and transmit power information of GB users, which it uses to determine the maximum amount of interference $\phi_m$ a GB user can tolerate and allocate it to the channel $m$ where the user can attain the same performance as in OMA. 
	\item It maps the threshold $\phi_m$ to each sub-channel, $\phi_m\xrightarrow{{\mathcal{M}}} m$.
	\item It broadcasts the $\phi = [\phi_1, \phi_2, \cdots \cdots, \phi_m]$ and the $\mathbf{ PP }$ corresponds to each sub-channel to GF users.
\end{itemize}

After receiving the above information from the BS via a broadcast signal, GF users randomly select a transmit power level independently from one of the available $\mathbf{ PP }$ and transmit on the sub-channel associated with that $\mathbf{ PP }$. Moreover, we assume that each GF user is allowed to select at most one sub-channel occupied by a GB user in a time slot $t_s$, as shown in Fig. \ref{PP}(\subref{p}). For this constraint, we define a sub-channel selection variable $k$, as follows:
\begin{align}\label{channel}
k_{m,j}{(t_s)}=
\begin{cases}
\text {1}, ~\text{ user $j\in  \mathbf{U}$ select subchannel $m$}\\~~~~~\text{which is occupied by GB user $i$}\; \\
\text{0}, ~~~\text{otherwise}
\end{cases}
\end{align}
The sum rate of the SGF-NOMA IoT networks can be maximized by selecting the sub-channel and optimal transmit power. Hence, by optimizing the transmit power $\mathbf P_t = \{P_1, P_2, \cdots, P_{NP}\}$ and sub-channel selection $\mathbf K_t = \{{k}_{1,1}, \cdots {k}_{m,j},  \cdots, {k}_{M,N_{GF}}\}$ for GF users, long-term throughput can be maximized. The data rate of a GB IoT user $i$ over sub-channel $m$ can be written as
\begin{align}\label{GB_rate}
R^{GB}_{m,i}{(t_s)}=B_s\log_2(1+\gamma^{GB}_{m,i}(t_s)),
\end{align}
whereas, the data rate of GF IoT user $j\in GF$ sharing the sub-channel $m$ with GB user $i$ can be expressed as
\begin{align}\label{GF_rate}
R^{GF}_{m,j}{(t_s)}=B_s\log_2(1+\gamma^{GF}_{m,j}(t_s)),
\end{align}
Based on (\ref{GB_rate}) and (\ref{GF_rate}) the cumulative capacity can be given as
\begin{align}\label{C_Data_Rate}
\bar{C}=\sum_{t_s=1}^{T_s} \sum_{m=1}^{M} \sum_{j=1}^{N_{GF,m}} R^{GF}_{m,j}{(t_s)} + \sum_{t_s=1}^{T_s} \sum_{m=1}^{M} \sum_{i=1}^{N_{GB,m}} R^{GB}_{m,i}{(t_s)}.
\end{align}

To determine the optimal transmit power levels for each cluster based PP, the optimization problem can be formulated as
\begin{subequations}\label{of}
	\begin{align}
	\underset{\mathbf{P}_t, \mathbf{K}_t}{\text{maximize}} ~~~~\bar{C}\\
	\textrm{s.t.}\quad
	P^{GB}_{m,1}h^{GB}_{m,1}(t_s) \ge P^{GF}_{m,1}&h^{GF}_{m,1}(t_s) \ge  \cdots \ge \nonumber\\ P^{GF}_{m,N_{GF,m}}h^{GF}_{m,N_{GF,m}}&(t_s),~\forall m, \forall t_s, \label{of0} \\
	\sum_{j=1}^{N_{GF,m}}P_{m,j\in \mathbf{U}}(t_s)~ \leq&~P_{max}, ~~\forall m, \forall t_s,\label{of1}\\
	\sum_{m=1}^{M} k_{i,j\in \mathbf{U}}(t_s) \leq 1~&, ~\forall j, \forall t_s, \label{of2}\\
	N_{G,m}(t_s)~~\geq~2, ~~&\forall m, \forall t_s, \label{of3}\\
	\sum_{m=1}^{M}R^{GB}_{m,i}{(t_s)} \geq \tau, ~&\forall i, \forall t_s, \label{of4}\\
	\sum_{m=1}^{M}R^{GF}_{m,j}{(t_s)} \geq \bar{\tau}, ~&\forall j, \forall t_s, \label{of5}\\
	\sum_{m=1}^{M}N_{GF,m}(t_s)\leq~& L_s,~~\forall t_s, \label{of6}
	\end{align}
\end{subequations}
where \eqref{of0} is the SIC decoding order and GB user data is decoding in first stage of SIC. Maximum transmit power limit of a user $j$ is given in \eqref{of1}. Constraint \eqref{of2} restricts the IoT users to select at most one sub-channel in a time slot $t_s$, \eqref{of3} represents the minimum number of IoT users to form a NOMA cluster.  \eqref{of4} is the required data rate of GB users to ensure QoS and \eqref{of5} represents the minimum required data rate threshold for GF users. \eqref{of6} shows the maximum number of GF users on each sub-channel.
\section{MA-DRL Framework for SGF-NOMA Systems}
DRL method aims to find good quality policies for decision-making problems and is able to evaluate the best utility among available actions with no prior information about the system model. DRL algorithms were originally proposed for a single agent interacting with a fully observable Markovian environment, with guaranteed convergence to the optimal solution. Recently, it has been widely used for more complex settings (e.g., MA-DRL (which is the extension of single-agent DRL)) and shows strong performance as compared to single-agent DRL algorithms \cite{8532121}.

\subsection{Modelling the Formulated Problem as a Stochastic Markov Game}
Stochastic games model dynamic interactions of the players (agents) where the environment changes in response to the player behaviour. Stochastic games move in stages, wherein each stage players select actions available to them in the current state. The chooses action has two effects: 1) They produce a stage reward; and 2) determine the next state probability. As a result, players (agents) receive a reward or penalty in the current state and try to receive high rewards in the next states. A Markov Game is an abstraction of the MDP \cite{sutton2018reinforcement}, MDPs are adopted in most modern RL problems that characterize the interaction between the environment and the agent. An MDP is a tuple of $(\mathbf{N}, \mathbf{S}, \mathbf{A},R(\cdot), \mathcal{P}(\cdot))$, where $\mathbf{N}$ is the number of agents, $\mathbf{S}$ is the set of states in the environment, $\mathbf{A}$ is the set of actions, which can be performed by an agent, $R(\cdot)$ is the immediate reward signal an agent receives from the environment for a given state-action pair, and $\mathcal{P}(\cdot)$ shows the transition probabilities between states. Agents act in the environment according to a certain policy $\pi(\cdot)$, which represents the probabilities that manage the agent's choice to perform an action based on the current state of the environment.
In RL algorithms, the agent's objective is to maximize their long-term rewards by making iterative variations to their policy as per the rewards they receive from the environment after taking actions. Briefly, these functions can be expressed mathematically as
\[\mathcal{P}(s, a, s' )=\mathbb{P}\big[S_{t+1}=s'| S_t =s, A_t=a \big],\]
\[R(s, a )=\mathbb{E}\big[R_{t+1} | S_t =s, A_t=a \big],\]
\[\pi(s, a )=\mathbb{P}\big[A_t=a  | S_t =s \big],\]
where $S_t$ represents the state of an agent at a learning step $t$ of an episode. The $A_t$ is the action the agent takes at that step. The $R_{t+1}$ is the reward received by the agent corresponding to the state-action pair.
\begin{figure*}[t!]
	\centering
	\includegraphics[width= 6.20 in, height= 3.70 in]{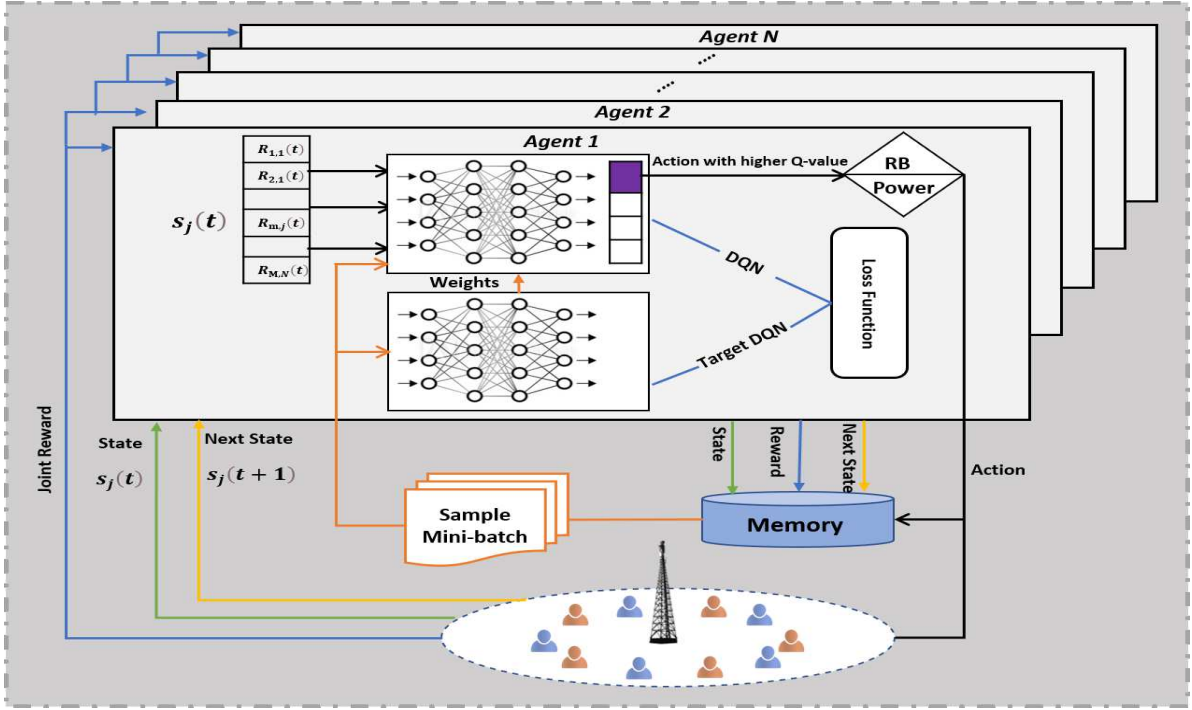}\\
	\caption{An illustrative structure of agents (with primary and target DQN) interacting with the wireless environment. Agents observe state form the environment and take joint action.}
	\label{MA-DRL}
\end{figure*}
Next, we show that resources selection (power and sub-channel) in SFG-NOMA systems can be modelled as an MDP. Detailed definitions are given below.
\begin{itemize}
	\item \textbf{Set of Agents} $\mathbf{N:}$
	In the SGF-NOMA scenario illustrated in Fig. \ref{system_model}, multiple GF users attempt to access the limited spectrum occupied by GB users. We define the GF IoT user\footnote{Transforming IoT devices into intelligent independent learners (agents) has two advantages: 1) (after training) It can reduce the frequent communication (e.g., CSI information) with the BS, which results in low signal overhead and reduce energy consumption by reducing control information exchange. 2) It will ease user collisions, which is a typical problem in GF communication and result in retransmissions and energy wastage.} 
	as an agent and interacts with the wireless communication environment as shown in Fig. \ref{MA-DRL}. Multiple IoT users jointly explore the environment and gain the experiences for their policy $\pi$ design. IoT users refine their policies based on the experiences gain from environment observations. All agents adjust their actions (sub-channel and transmit power selection) to obtain an optimal policy $\pi^*$ by maximizing the reward \cite{8792382}. 
	At each time-step (TS) $t$, each agent $j$ receive a state
	$s_j(t)$, and performs an action $a_j(t)$ that forms a joint action as
	\begin{align}
		{a}(t)=(a_1(t), a_2(t), \cdots, a_j(t), \cdots, a_N(t)).
	\end{align}
	Thereafter, the environment return a reward $r_j(t+1)$ and next state to the current agent.
	\item {\textbf{State space} $\mathbf{S:}$ All agents collectively explore the wireless environment by observing different states from the environment. More specifically, we represent the sum-rate of GF users as current state $s_j(t) \in S_j$ in time step $t$
	} 
	\begin{align}\label{state}
	S_j= \{R^{GF}_{1,1}(t),R^{GF}_{2,1}(t),\dotsi,R^{GF}_{M,N_{GF}}(t)\},
	\end{align}
	where $R^{GF}_{m,j}$ is the data rate of GF user $j$ on sub-channel $m$ and depends on previous time slot $(t-1)$. 
	\begin{remark}
	After decoding all users' information, the BS learns data rates of GF users. As a result, the state $s(t)$  and reward $r(t)$ is available at the BS in each time slot. Therefore, providing data rates of users as a state to the agents can prevent direct collaboration among the agents which can further reduce the signalling overhead and energy consumption.
	\end{remark}
	
	\item \textbf{Action Space} $\mathbf{A:}$ 
	We define the action of a GF user $j$ as a selection of transmit power and sub-channel. The action space of agent $j$ can be expressed as
	\begin{align}\label{action}
	A_j(t)= \{1,2,\cdot\cdot\cdot pm,\cdot\cdot\cdot,P_{NP}M\}.
	\end{align}
	We use a set of discrete transmit power levels $\textbf{P}_t=\{P_1, P_2, \cdots, P_{NP}\}$. Agent $j$ is only allowed to select one power level in time slot $t$ to update its transmit power strategy. All agents have same action space $[A_1= A_2= \cdots = A_j= \cdots= A_N]$ for $\forall j \in \mathbf{N}$. Then the combination of action spaces of all agents can be expressed as
	\begin{align}
		A =(A_1\times A_2 \times \cdots \times A_j \cdots \times A_N).
	\end{align}
	 Action space dimension is $M\times P_{NP}$, where $M$ is the number of sub-channels and $P_{NP}$ is the number of available discrete transmit power levels.
	\item \textbf{Reward Engineering} $Re:$ The reward function classifies the actions of an agent as good or bad. Therefore the reward function must be designed in such a way that correlates with the objective function. In the SGF-NOMA environment, each agent operates in a network-centred manner and receives network throughput as a reward from the BS via a broadcast signal.  Markov Games, in which all agents receive the same reward, is known as Team Markov Games (TMGs) \cite{nowe2012game}. The reward function of each agent $j\in \mathbf{N}$ at learning step $t$ is given as
	\begin{multline}\label{Re}
	r_j(t)=
	\begin{cases}
	\text {$ \bar{C}$}, \text{ if $\bar{C}{(t+1)}\geq \bar{C}{(t)}$ } \text{and ensure }\\
	~~~~\text{constraints given in \eqref{of0}-\eqref{of6}},\\
	\text{0},  ~~\text{otherwise}.
	\end{cases}
	\end{multline}
\item \textbf{Transition Probability} $\mathcal{P:}$ The state transition probability function is the probability of transitioning to next state $s(t+1)$ after taking a joint action $a(t)$ in current state $s(t)$.
\end{itemize}

We define Q function $Q_j^\pi(s_j(t), a_j(t))$ associated with policy $\pi$ as the expected cumulative discounted reward for each agent $j$ after taking action $a_j$ in state $s_j$ i.e.,
\begin{align}\label{Q-func}
Q_j^\pi(s_j, a_j) = \mathbb E^\pi\Big[Re{(t)}\big|s_j{(t)}=s, a_j{(t)}=a\Big],
\end{align}
where $Re$ is the long-term accumulated and discounted reward and calculated as
\begin{align}\label{Reward1}
Re = \sum_{k=0}^{K}\beta^k r^{(t+k+1)}, ~~~~~ 0<\beta \leq 1,
\end{align}
where $\beta$, $k$ and $K$ represents the discount factor, epoch and maximum epoch respectively. The policy $\pi$ map the state $s(t)$ to corresponding Q-value under action $a(t)$. If an agent $j$, observe a state $s_j{(t)}$, perform action $a_j{(t)}$, receiving a reward $r{(t)}$ and the next state $s_j{(t +1)}$, then its corresponding Q-value in the Q-table is updated as
\begin{align}\label{Qup}
Q(s_j{(t)}, a_j{(t)}) &\leftarrow r{(t)}+\beta \max_{a_j\in {A}_j} Q(s_j{(t+1)}, a_j).
\end{align}
All agents aim to maximize the expected reward that leads the agents to derive an optimal policy $\pi^*$. Once the optimal Q function $Q^*(s, a)$ is obtained, every agent finds an optimal policy $\pi^*$ such that $Q^*(s, a) \geq Q(s, a)~\forall s\in S ~\text{and}~ a\in A$. In a stochastic TMG, the joint optimal policy is known as Nash equilibrium (NE) and can be described as $\pi^*=(\pi_1^*, \pi_2^*, \cdots, \pi_N^*)$. 
The NE is a combination of policies of all agents, and each policy is the best retaliation to other policies. Therefore, in a NE, each agent's action is the best reaction to the other agent's action choice. More specifically, an agent's strategy or policy must be consistent with the policies of other agents, otherwise, no agent can gain an advantage by changing its strategy \cite{nasir2019multi}. Hence, each learning agent job is to explore a NE for any state of the environment. The classic Q learning \cite{sutton2018reinforcement} maintain a Q-table to record Q-values for each state-action pair. However, for the IoT scenario,  Q-table size increases with the increasing state-action spaces (an increase of IoT users) that makes Q-learning expensive in terms of memory and computation. Therefore, the deep Q learning \cite{mnih2015human} is proposed to overcome the aforementioned problem by combining the Q learning with Deep Neural Network (DNN) with weights $\theta$ for Q function approximation $Q(s, a; \theta)$. In MA-DRL, each agent consists of a primary (online) network, target network (both networks with the same architecture) and a replay memory. During the training phase of DNN, the learnable parameters (weights and biases) are updated according to system transition history consists of a tuple of $\big(s_j{(t)}, a_j{(t)}, r{(t)}, s_j{(t+1)}\big)$ that enhance the accuracy of Q-function approximation. In a learning step $t$, each agent $j$ input the current state $s_j$ to the DQN and output Q-values corresponding to all actions. The agent select action with the highest Q-value and obtain an experience in the form of a tuple $\big(s_j{(t)}, a_j{(t)}, r{(t)}, s_j{(t+1)}\big)$ and store it to replay memory. To update the weights $\theta$ of the target Q-network, a mini-batch of data is randomly sampled from the replay memory. The target value produced by target Q-network from randomly sampled tuple is given as
\begin{align}\label{argmax1}
y_j{(t)} & =r{(t)}+ \beta\argmax_{a_j{(t+1)}\in A_j} \;Q(s_j{(t+1)}, a_j{(t+1)}; \theta).
\end{align}
\begin{figure*}[t!]
	\centering
	\begin{subfigure}{.5\linewidth}
		\centering
		\includegraphics[height= 1.90in, width=3.50in]{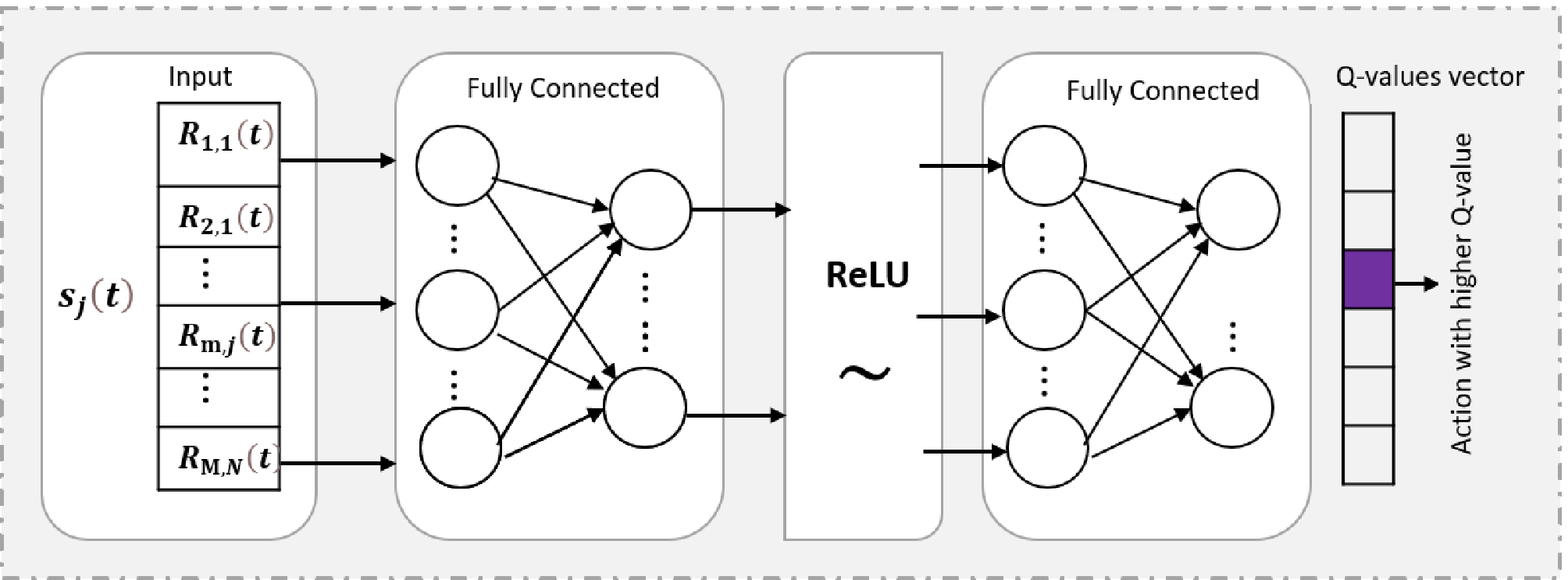}
		\caption{DNN structure for DDQN}
		\label{DQN}
	\end{subfigure}%
	\begin{subfigure}{.5\linewidth}
		\centering
		\includegraphics[height=1.90in, width= 3.50in]{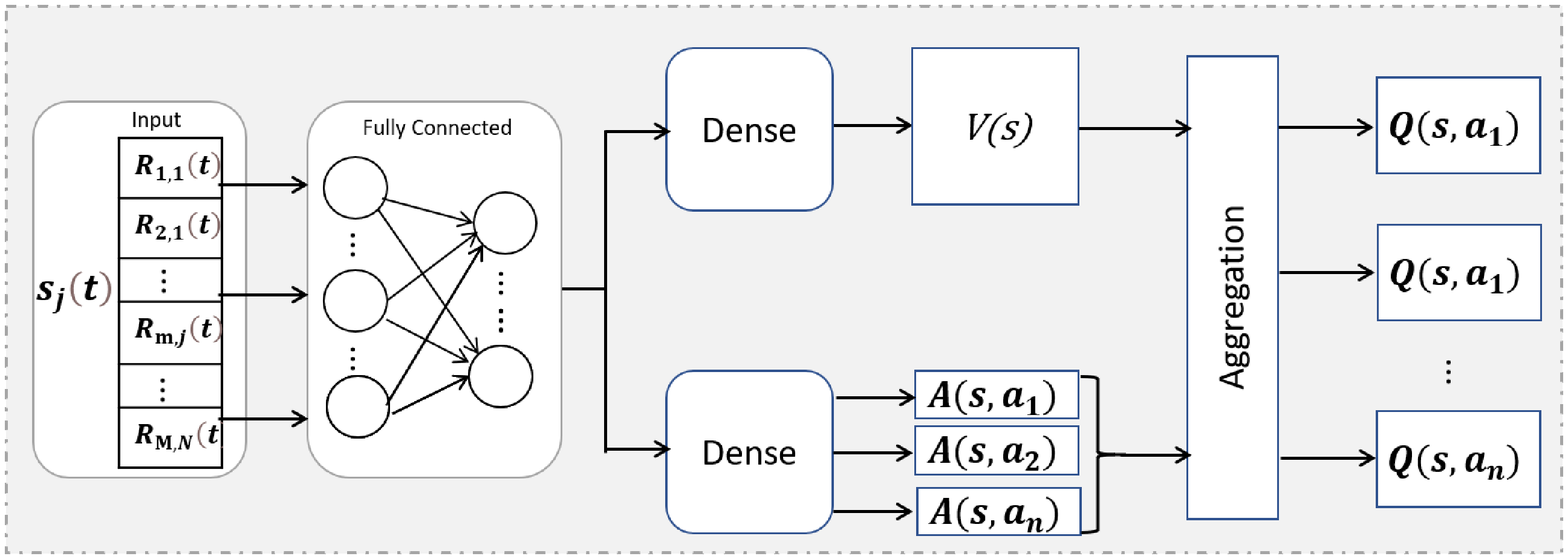}
		 \caption{DNN structure for Dueling DDQN}
		\label{DQN1}
	\end{subfigure}%
	\caption{DNN structures of the proposed algorithms: (a) DNN structure for DDQN algorithm, which consist of input layer followed by hidden layers and an output layer. (b) DNN structure for Dueling DDQN algorithm, which consist of input layer followed by fully connected layer and dueling layer.}
	\label{str}
\end{figure*}
Agents in DQN use the same Q-values for action selection and action evaluation, which leads to a Q-value overestimation problem and algorithm converging with a non-optimal solution because for action selection and action evaluation the max operator use the same value. To mitigate this problem and improve agents learning efficiency, the following two improved versions of DQN are proposed.

\subsection{Double Deep Q-Network Algorithm}
The DDQN \cite{van2015deep} prevents the above mentioned problem by decoupling the max operation in the target network in action selection and action evaluation. More specifically, we use two neural networks (NNs) DQN$_1$ and DQN$_2$, where DQN$_1$ is used to select actions and DQN$_2$ is used to evaluate the corresponding Q-value of that action. For DDQN, the target of DQN can be rewritten as:
\begin{multline}\label{DDQNV}
y_j{(t)}  =r{(t)}+ \beta Q(s_j{(t+1)}, \\\argmax_{a_j{(t+1)}\in A_j} \;Q(s_j{(t+1)}, a_j{(t+1)}; \theta); \theta).
\end{multline}
Using a variant of stochastic gradient descent (SGD) \cite{lecun2015deep}, the primary Q-network can be trained by minimizing the loss function. 
\begin{align}\label{loss}
Loss(\theta) & = (y_j{(t)}-Q_j{(t)}(s_j{(t)}, a_j{(t)}; \theta))^2.
\end{align}
The proposed MA-DDQN algorithm for transmit power and sub-channel selection with proposed DNN architecture depicted in Fig. \ref{str}(\subref{DQN}) is given in \textbf{Algorithm 1}.
\begin{algorithm}
	\small
	\caption{Proposed MA-DDQN based SGF-NOMA Algorithm}
	\begin{algorithmic}[1]
		\State Initialize primary network with random weights $\theta$ 
		\State  initialize target Q-network with same weights as primary network
		\State Initialize replay memory with size $Z$, and other training parameters $\beta$, $\epsilon$
		\For {episode = 1 to M}
		\State reset initial state of the environment
		\For {time-step = 1 to N}
		\State Input state $s(t)$
		\State Take joint action $a(t)$ following $\epsilon- greedy$  policy, receive next state $s(t+1)$ and reward $r(t)$		
		\State Store  $s(t), a(t), r(t), s(t+1)$ in replay memory $Z$
		\State Sample mini-batches from memory $Z$
		\State minimize the loss between the primary network and target network using \eqref{loss}
		\If {  episode\% ==$U_{steps}$ }
		\State copy primary network weights to target Q-network weights
		\EndIf
		\EndFor
		\EndFor
	\end{algorithmic}
\end{algorithm}
\begin{algorithm}
	\small
	\caption{Proposed MA-Dueling DDQN based SGF-NOMA Algorithm}
	\begin{algorithmic}[1]
		\State Repeating lines 1-10 in \textbf{Algorithm 1}
		\State Calculate two streams of the evaluated deep network  $V_\pi(s;\theta, \theta^V)$ and $A_\pi(s,a;\theta, \theta^A)$,  and combine them using (\ref{duq})
		\State Minimize the loss between the primary and target network
		\If {  episode\% ==$U_{steps}$ }
		\State copy primary network weights to target Q-network weights
		\EndIf
	\end{algorithmic}
\end{algorithm}
\subsection{Dueling Double Deep Q Network Algorithm }
The conventional DQN algorithm computes the value of each action in a specific state. However, different policies may result in the same value function in some states. This phenomenon may slow down the learning process in identifying the best action in a given state. Therefore, the dueling DDQN can generalize the learning process for all actions and has the ability to identify the best actions and important states quickly without learning the effects of each action for each state. The dueling DQN \cite{wang2016dueling} is the improved version of DQN, where the Q-network has two streams (sequences) Q-function (i.e., the state-action value function is decomposed), namely the state value function $V_\pi (s)$ and the advantage function $A_\pi(s, a)$, to speed up the convergence and improve the efficiency. The value function $V_\pi (s)$ is used to represent the quality of being in a particular state (calculating the average contribution of a particular state to the Q-function), and the advantage function $A_\pi(s, a)$ measures the comparative importance of a particular action versus other actions in a particular state. Output of the dueling network is obtained by combining these two streams to form an aggregate module and a single output Q-function as follows:
\begin{align}\label{duq}
	Q_\pi(s,a; \theta, \theta^V, \theta^A) = V_\pi(s;\theta, \theta^V) + A_\pi(s,a;\theta, \theta^A),
\end{align}
where, $\theta, \theta^V, \theta^A$ represents parameters of common network, value stream parameters and advantage stream parameters, respectively. Practically, the agent cannot distinguish between $V_\pi(s) $ and $A_\pi(s,a)$. Since the agent may not be able to obtain a unique solution for $Q_\pi(s,a; \theta, \theta^V, \theta^A)$, it may be unidentifiable and result in poor performance. To solve this problem, the Q-values for each action $a$ in state $s$ are generated by the aggregation layer as follows:
 \begin{multline}\label{du}
 Q_\pi(s,a; \theta, \theta^V, \theta^A) = V_\pi(s;\theta, \theta^V) + A_\pi(s,a;\theta, \theta^A)- \\\frac{1}{|\mathcal{A}|} \sum_{a(t+1)}A_\pi(s(t),a(t+1);\theta, \theta^A).
 \end{multline}
The operation of (\ref{du}) guarantees that the dominant function of each action in this state remains unchanged and reduces the range of Q-value and excess degrees of freedom, which improves stability. The proposed MA-Dueling DDQN algorithm for transmit power and sub-channel selection with the proposed DNN architecture shown in Fig. \ref{str}(\subref{DQN1}) is given in \textbf{Algorithm 2}.

\subsection{Proposed MA-SGF-NOMA Algorithm}
In our proposed MA-DRL algorithms, each GF user acts as an agent and runs an independent DQN. All agents collectively explore the wireless environment and learn an optimal policy to find a NE. For exploration and exploitation trade-off, we use the $\epsilon-greedy$ method. To fully explore the environment and find the action with the best reward, the agent considers a random action with probability $\epsilon \in [0,1]$. Whereas to exploit better actions, the GF user selects the best action associated with the largest Q-value in a given state with probability 1$-\epsilon$. In a single learning step $t$, each GF user $j$ uploads the current state $s_j(t)$ to its primary Q-network and acquire all the Q-values that correlates to all actions. The agent then decides its action according to the $\epsilon-greedy$ method and takes joint action $a(t)$. The environment transitions to a new state $s(t+1)$ with probability $\mathcal{P}$ and all agents (GF users) receives the same reward (system throughput). In each step $t$, agents forms a new experience by interacting with the wireless environment and stores it to memory $Z$ in the form of a tuple $(s(t), a(t), r(t), s(t+1))$. To compute the Q-value of the target network, we randomly sample mini-batches of stored transitions from the replay memory. In each training iteration, to improve the policy $\pi$, the primary Q-network is trained by minimizing the gap (error) between the actual value and predicted value with the SGD method using (\ref{loss}). After a fixed number of training iterations, primary network weights are copied to the target Q-network. At the end of the training process, each agent $j$ finds an optimal policy $\pi^*_j$ that forms global (joint) optimal policy $\pi^*$.
	\begin{center}
	\begin{table*}[!]
		\centering
		\scriptsize
		\caption{Quantitative comparison of the signalling overhead}
		\label{tab2}
		\begin{tabular}{ |c|c|c|c| } 
			\hline
			Reference & Overhead for $U$ GF and $V$ GB users & Power decision & Optimization method \\ \hline
			\cite{8662677} & $\big( \underbrace{4V}_{\text{power}}+ \underbrace{2V}_{\text{CSI}}+ \underbrace{2}_{\text{channel quality threshold}} \text{or} \underbrace{2U}_{\text{beacon transmission}}\big)$ bits & BS & Conventional \\  
			\hline
			\cite{ding2020new} & $\big( \underbrace{4V}_{\text{power}}+ \underbrace{2V}_{\text{CSI}}+ \underbrace{2}_{\text{data rate threshold}} \big)$ bits & BS & Conventional \\  
			\hline
			\cite{lu2020advanced} & $\big(\underbrace{2}_{\text{pilot}} + \underbrace{2V}_{\text{SNR}}+ \underbrace{2V}_{\text{CSI}}+ \underbrace{2}_{\text{target rate}}+ \underbrace{2}_{\text{decoding threshold}}+ \underbrace{2}_{\text{SNR threshold}}\big)$ bits & BS & Conventional \\  
			\hline
			\cite{liu2021rate} & $\big(\underbrace{2}_{\text{pilot}} + \underbrace{4V}_{\text{power}}+ \underbrace{2V}_{\text{CSI}}+ \underbrace{2}_{\text{interference threshold}}\big)$ bits & BS & Conventional \\  
			\hline
			\cite{9686696} & $\big(\underbrace{10U}_{\text{state}} + \underbrace{2}_{\text{reward}}\big)$ bits & User & RL based \\ \hline
			Proposed & $\big(\underbrace{2U}_{\text{state}} + \underbrace{2}_{\text{reward}}+ \underbrace{16}_{\text{PPs}}\big)$ bits & User & RL based \\  \hline
		\end{tabular}
	\end{table*}
\end{center}
{\subsection{Analysis of the Proposed Algorithm}
	\begin{enumerate}
\item Computational Complexity Analysis:
The computational complexity of DRL algorithms is assessed from two perspectives: predicting and model training (expensive). Assume that the DNN in use has $L$ layers, each layer with $g_l$ neural nodes, and the input layer is of size X, which is equal to the number of active GF users. The estimated complexity to complete a single prediction (predicting the power and sub-channel selection policies) is given by $\mathcal{O}(Xg_1+\sum_{l=1}^{L}g_lg_{l+1})$. The training computational complexity of the proposed algorithm until convergence is of order $\mathcal{O}(N_t  M  N  (Xg_1+\sum_{l=1}^{L}g_lg_{l+1}))$, where $N_t$ is the total number of agents, $M$ is the number of episodes and $N$ is the learning steps. 
	\item Convergence Analysis:
The convergence of a multi-agent system refers to whether the agents' combined strategy finally approaches to a NE in response to joint actions. In our algorithm, we update the weights $\theta_j$ for the $Q_j$ network by adopting gradient descent method, in which the learning rate decreases exponentially with iterations. The weight $\theta_j$ will therefore converge to a certain value after a finite number of iterations, ensuring the convergence of the proposed algorithm. In fact, analysing the convergence of neural networks theoretically before training is difficult, as mentioned in \cite{8654727}. The analytical difficulty arises from the fact that the convergence of a neural network is strongly dependent on the hyperparameters utilised during the training process \cite{9058679}. In this context, there is a complex quantitative relationship between network convergence and hyperparameters. Instead, we use simulation to demonstrate the convergence of our approach.
\item Signalling overhead Analysis: The overhead is determined by the number of information bits required to feed back sub-channel indicators, channel status data, and a specific user's transmission power over a sub-channel \cite{nouruzi2022toward}. Additionally, in ML based approaches, the exchange of states and rewards between the agent and environment also affect the overhead. 
Similar to \cite{nouruzi2022toward}, we assume the set \{16, 4, 4\} as the number of information bits to transmit channel status, sub-channel indicators, and the transmission power in the feedback process and 2 bits for obtaining a single value of a state and reward. The conventional optimization methods produce high overhead because mostly depends on instantaneous CSIs and other threshold information. Moreover, the power decision in these methods is centralized and the complexity is located at the BS. Therefore, users have to adjust their transmission powers based on what the BS sends back. The ML based approach given in \cite{9686696} also leads to large overhead because the environment states consists of three values, i.e., channel gain, transmit power and sub-channel indicator. A quantitative comparison in terms of signalling overhead is given in table \ref{tab2}.  
\end{enumerate}

\section{Numericals Results}
\subsection{Simulation Setup and Training Parameters}
In this section, we evaluate the performance results of our proposed scheme. BS is located at the centre of a circle with a radius of $1000m$. The GF and GB users follow a Poisson distribution throughout the cell area. We set the path-loss exponent $\alpha=3.0$, $n_0= -90$~dBm and the sub-channel bandwidth is $10$~KHz \cite{zhang2020deep}. In addition, the GB users transmit data at fixed power, while GF users select the transmit power from the available power levels $\{0.1, 0.2, 0.3, 0.4, 0.5, 0.6, 0.7,$ $0.8, 0.9\}$~w, and choose among the $3$ sub-channels occupied by GB users. To ensure the QoS of GB users, we set a threshold data rate (spectral efficiency) $\tau = 15$~bps/Hz. A minimum rate of 4 bps/Hz is required for GF users. Next, we define the hyperparameters used for the architecture
of our DQN algorithm.  Our proposed algorithm trains a DQN with three hidden layers have $X_1$ = 250, $X_2$ = 120, and $X_3$ = 60 neurons, respectively. We use this relatively small network architecture for training purpose, to ensure that agent decides their decisions (actions) as fast as possible. We use Rectified Linear Unit (ReLU) as an activation function and Adam optimizer to increase the learning rate for fast convergence. We set the learning rate to 0.001 and discount factor $\beta$ = 0.9\cite{zhang2020deep}. 
We set the memory size $Z=10000$, batch size of 32 and target network update frequency to 1000.
Additionally, initially, the $\epsilon$ value is set to 1.0 and gradually decreases to a final value of 0.01 to balance the exploration and exploitation phenomena. The training lasts for 500 episodes
and each episode consists of 100 time steps.

\begin{figure*}[!h]
	\centering
	\begin{subfigure}{.5\linewidth}
		\centering
		\includegraphics[height= 2.2in, width=3.3in]{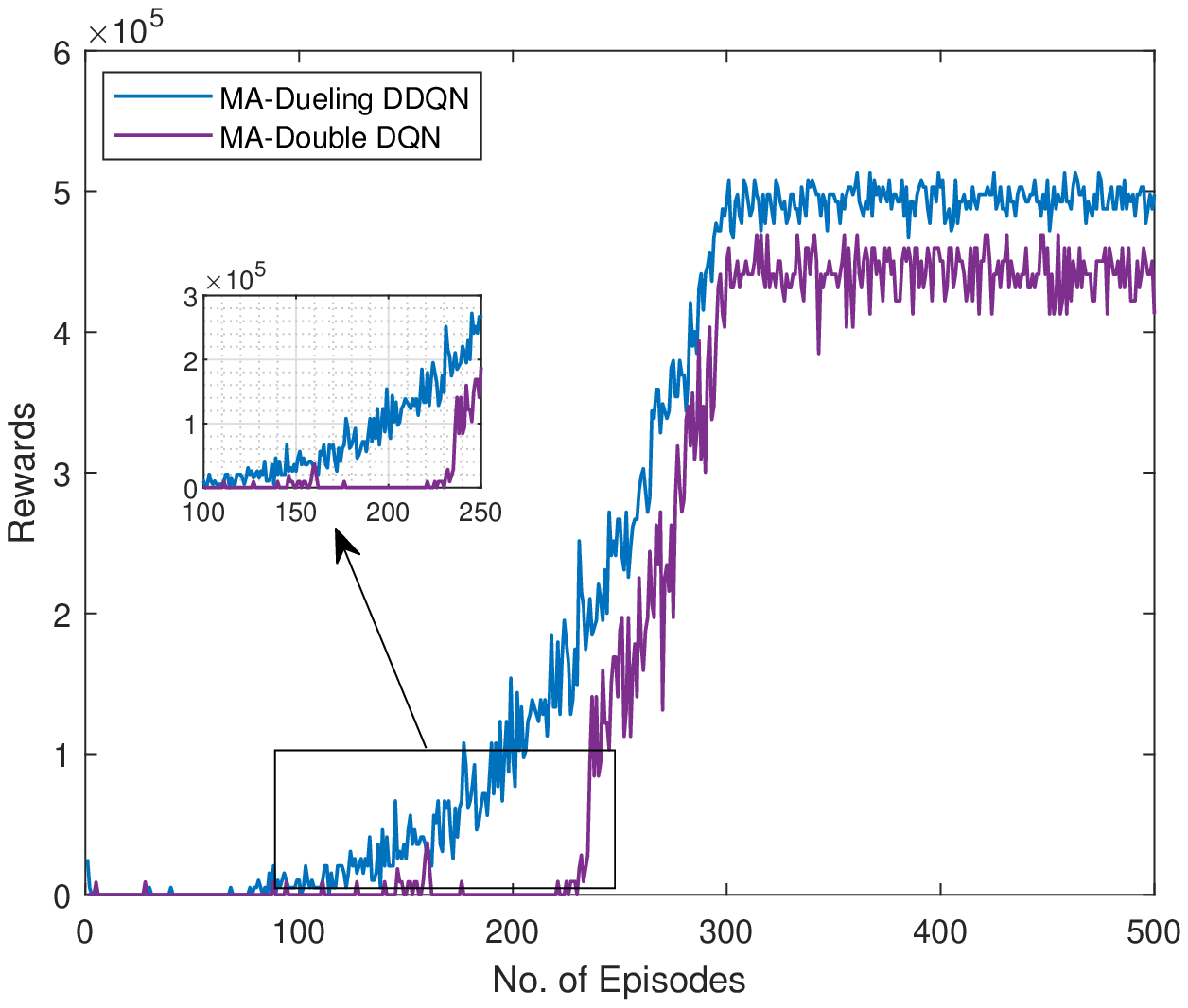}
		\caption{Performance w.r.t. large state and action set}
		\label{large}
	\end{subfigure}%
	\begin{subfigure}{.5\linewidth}
		\centering
		\includegraphics[height= 2.2in, width=3.3in]{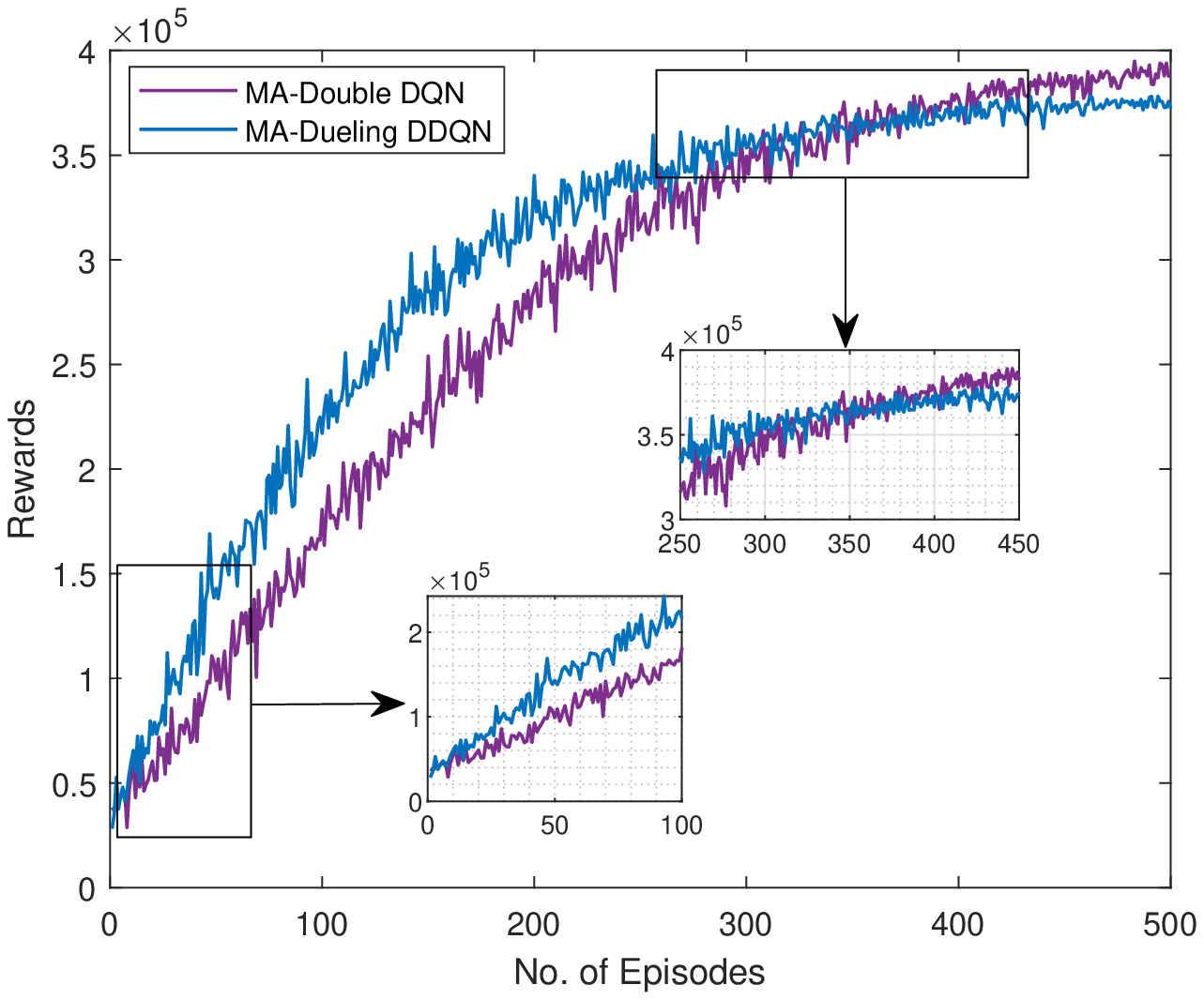}
		\caption{Performance w.r.t. small state and action set}
		\label{small}
	\end{subfigure}\\
	\begin{subfigure}{.5\linewidth}
		\centering
		\includegraphics[height= 2.2in, width=3.3in]{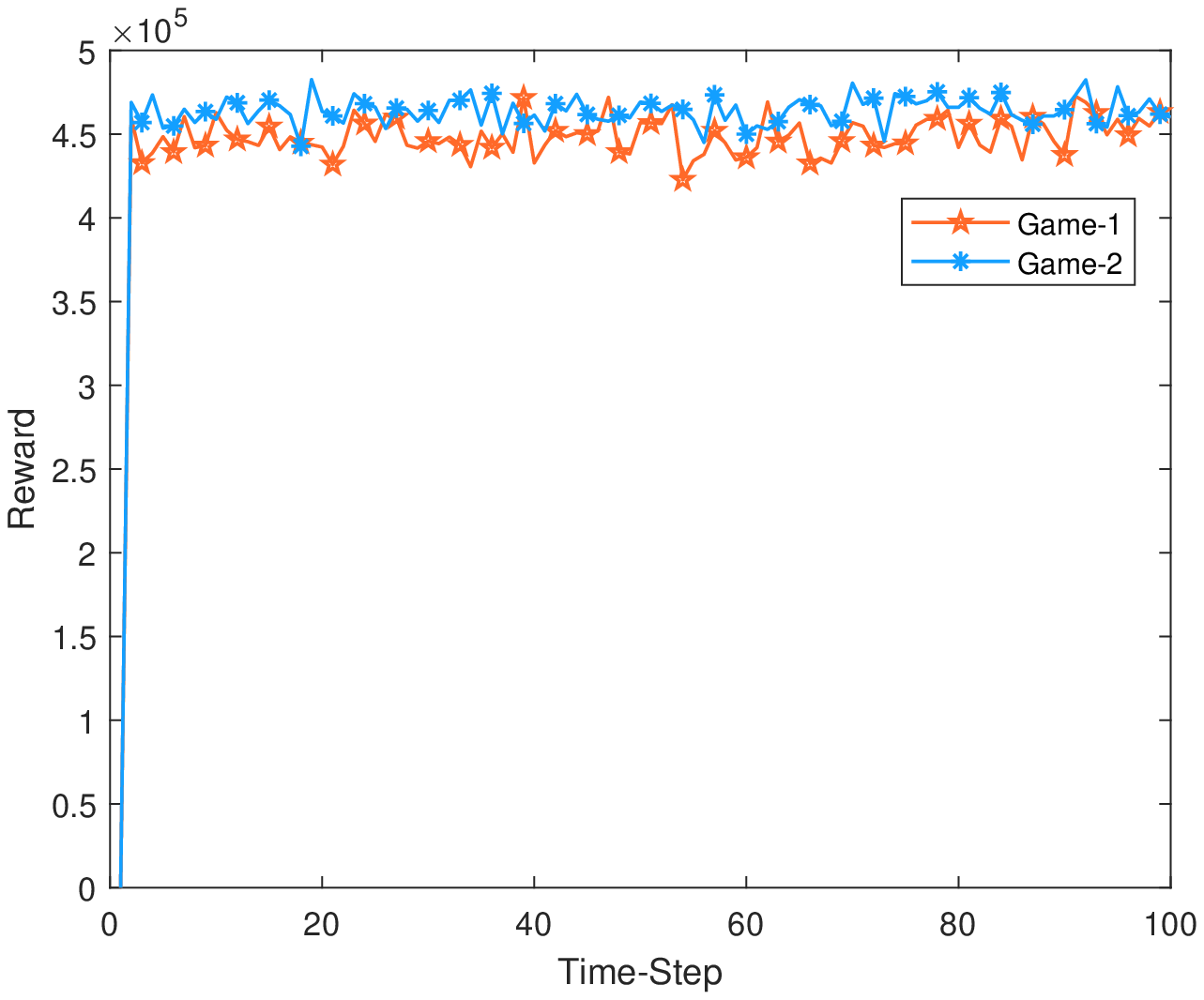}
		\caption{Required time steps vs. convergence}
		\label{ts}
	\end{subfigure}%
	\begin{subfigure}{.5\linewidth}
		\centering
		\includegraphics[height= 2.2in, width=3.3in]{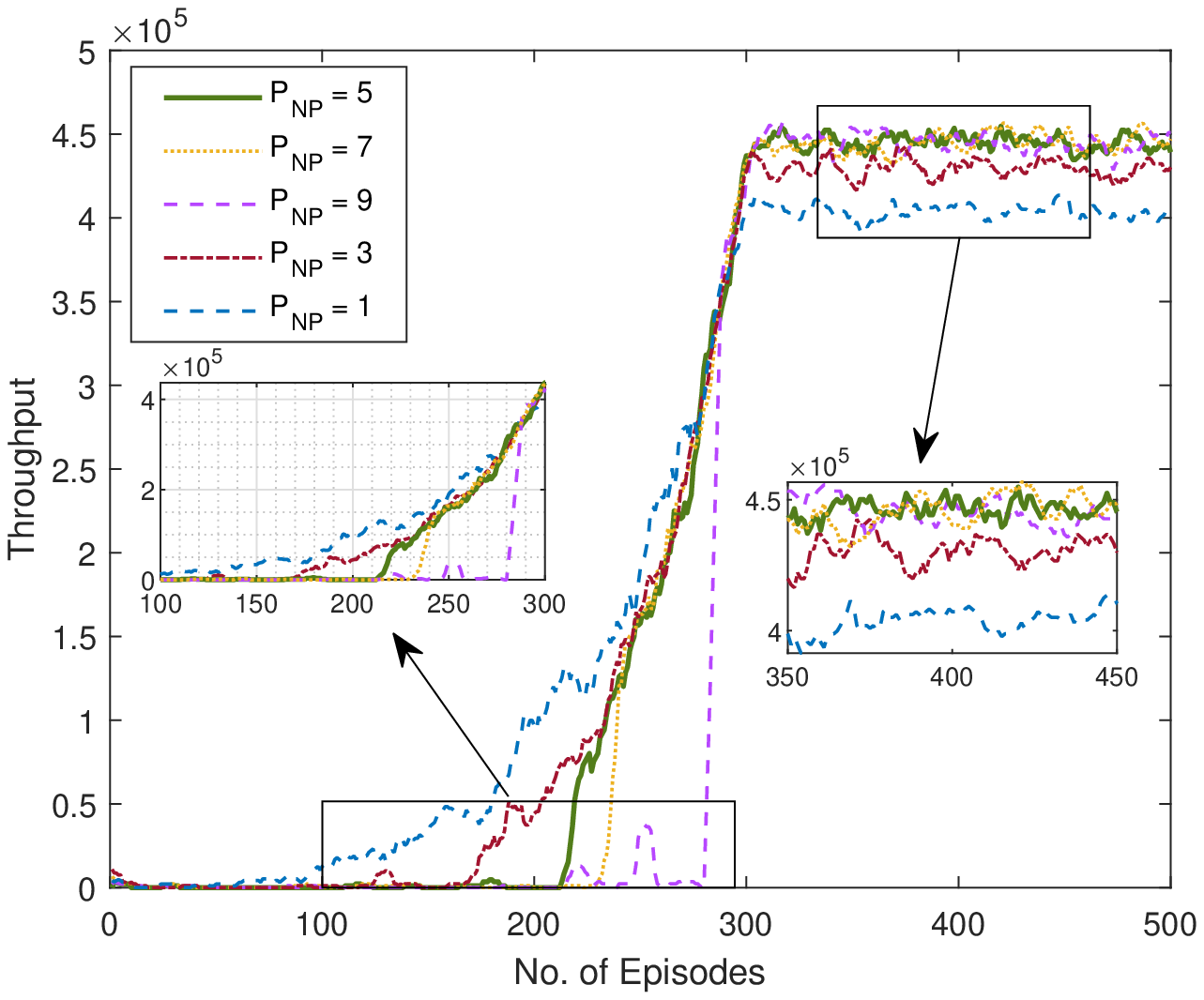}
		\caption{No. of Power Levels }
		\label{PL}
	\end{subfigure}%
	\caption{ The convergence comparison: Sub-figure (a) represents the convergence comparison against large action and state spaces. Sub-figure (b) shows the performance against small action and state spaces. Sub-figure (c) represents the time step required for convergence. Sub-figure (d) shows network throughput w.r.t. different number of power levels.}
	\label{conv}
\end{figure*}
\subsection{Learning Performance and Convergence Analysis}

Fig. \ref{conv}(\subref{large}) shows the learning efficiency and convergence comparison of MA-DDQN and MA-dueling DDQN based SGF-NOMA algorithms against large number of action and state spaces. The worst performance of both algorithms is seen at the beginning due to random action selection in the exploration phase. However, after gaining some experience by interacting with the wireless environment, MA-Dueling DDQN has better learning performance as compared to the MA-DDQN algorithm. Agents in MA-dueling DDQN algorithm learn and refine their policies faster and start converging after about 100 episodes. On the other hand, agents in the MA-DDQN algorithm learn slowly and start converging after 220 episodes. This is because different policies may lead to the same value function in some states and this phenomenon slow down the learning process in identifying the best action in a particular state. However, the MA-Dueling DDQN generalizes the learning process for all actions and can quickly identify the best actions and important states without learning the effects of each action for each state, which speeds up the learning process for each agent. During the exploitation phase, the agents in both algorithms exploit the environment with better actions and gradually increase the reward value and reach its maximum in 300 episodes. However, the MA-Dueling DDQN shows fast learning efficiency and better performance in terms of rewards (system throughput) gain. Fig. \ref{conv}(\subref{small}) shows the learning behaviour of both algorithms against a relatively small set of actions and states. Both algorithms perform almost in the same manner in terms of learning and rewards gain. It can be concluded that MA-DDQN performs better on problems with limited action space. Therefore, MA-DDQN can be used for problems with a small set of actions instead of MA-Dueling DDQN, which requires training a separate neural network to compute the function estimator. Moreover, identifying the best actions and important states is only desired in large action and state spaces to improve the learning process, for which MA-Dueling DDQN is the best option.
Fig. \ref{conv}(\subref{ts}) illustrates the inference of the training policy for a single episode in two different games. During the inference (using no value function), agents select the action with the highest Q value. For different channel realization and random initialization, the algorithm converges within 2 time steps for both games.
	
\subsection{Impact of Number of Power Levels}
In this section, we discretize the transmit power into different power levels to evaluate the network performance with varying number of power levels. Fig. \ref{conv}(\subref{PL}) shows the impact of the number of power levels on network performance in terms of throughput and convergence. Small number of power levels reduces the action space $(M\times P_{NP}, (3\times 1 = 3))$ which results in quick convergence. It can be observed that after about 100 episodes, each agent (GF users) find an optimal policy and receive a continuous reward in terms of throughput. However, it results in the lowest network throughput because each GF user has a limited choice of transmit power levels. Network with $P_{NP} = 3$ increases the action space from to 3 to 9 (i.e., $3\times3 = 9$), which required more training episodes to find an optimal policy. From the figure, it can be seen that after 175 episodes agent's receiving a reward in terms of throughput. With more transmit power levels, the throughput of the network increases since each user selects transmit power fairly according to the channel gain that reduces interference to other users in the cluster. Next, we evaluate network performance w.r.t. $P_{NP} = 5$, which provides the best performance results compared to the rest of the power levels. The throughput increases continuously from 200 episodes reaches to peak throughput in 300 episodes and remain converge till the end of the training. Algorithm with $P_{NP} = 7$ and $P_{NP}= 9$, users spends longer training time to explore the environment with best actions and provides throughput same as $P_{NP}= 5$. With more power levels, most of the actions (with high transmit power level) becomes invalid due to users transmit power constraints in each sub-channel. Thus, increasing the number of power levels does not always help improve system performance, and it becomes difficult to figure out the best actions in large action space.
\begin{figure}
	\centering
	\begin{subfigure}[b]{0.5\textwidth}
		\includegraphics[width=1\linewidth]{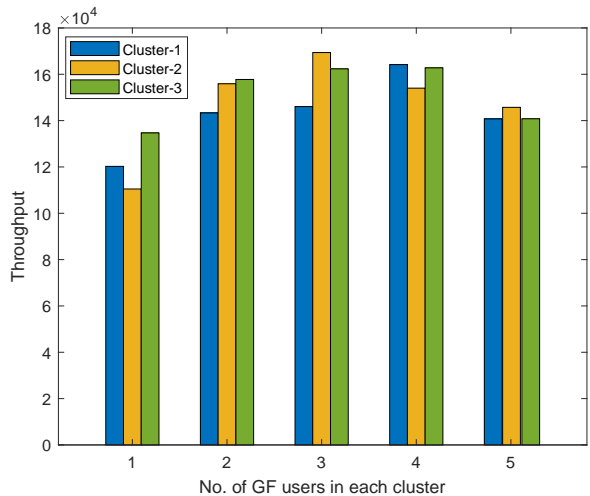}
		\caption{Clusters throughput}
		\label{cluster} 
	\end{subfigure}
	
	\begin{subfigure}[b]{0.5\textwidth}
		\includegraphics[width=1\linewidth]{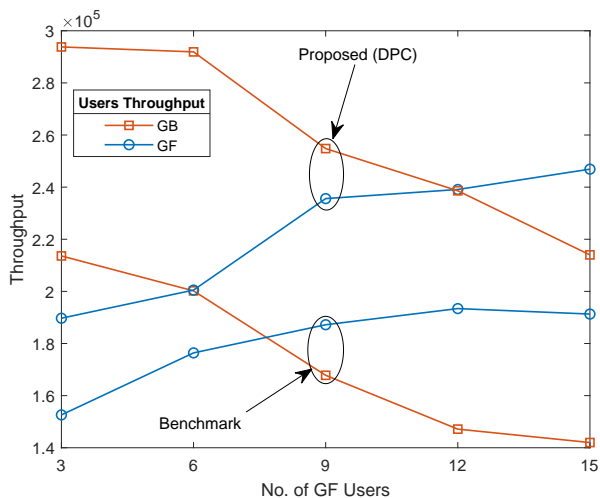}
		\caption{ GF vs GB users throughput}
		\label{dpc}
	\end{subfigure}
	
	\caption[Two numerical solutions]{Cluster throughput, GB and GF users throughput: Sub-figure (a) represents the throughput of each cluster in the network of our proposed algorithm. Sub-figure (b) Shows the impact of the PC strategy on the throughput of GB and GF users in the proposed algorithm and benchmark scheme \cite{8662677}.}
	\label{pol}
\end{figure}
\subsection{Impact of Number of Users in a Cluster}
We consider the number of GB and GF users on each sub-channel as a cluster. To design the PP for each sub-channel, we first determine the upper bound of GF users in each cluster to maximize the network throughput while maintaining the QoS of GB users. Fig. \ref{pol}(\subref{cluster}) shows the impact of an increasing number of GF users on cluster throughput. The throughput of each cluster has been increased with the increasing number of GF users on each cluster till to serve at most 4 GF users on each cluster (sub-channel). The proposed algorithm can assign the optimal transmit power level to the varying number of GF users that increases clusters throughput. Serving more than four users on each cluster decreases cluster throughput due to increasing interference and limiting the transmit power due to transmit power constraint. More specifically, GF users are restricted to transmit with the lowest power level to ensure the QoS of GB users that declined throughput of each cluster. Further increasing the number of GF users on each cluster results in the lowest throughput and leads to SIC and decoding failure at the BS due to severe interference. It is worth noting that multiplexing large number of users on each sub-channel may impose extra latency due to SIC processes. Moreover, to ensure the QoS of GB users, we utilize a fixed (open loop) QoS threshold. 
Investigating an adaptive QoS threshold for both GF and GB users can further improve clusters throughput.

\subsection{PC Impact on GF and GB Users Performance}
	The SGF-NOMA improves connectivity by allocating the additional data rate of GB users to GF users. However, admitting GF users to the same sub-channel occupied by GB users causes interference to the GB users. Therefore, the transmit power selection mechanism has a strong correlation with the throughput of both types of users. Fig. \ref{pol}(\subref{dpc}) shows how the power allocation mechanism affects the throughput of GF and GB users. The throughput of GF users increases with DPC (proposed), while that of GB users decreases.  More specifically, GF users as an agent select those power levels that cause limited interference to GB users and maximize their throughput.  The GF users in the benchmark scheme, transmit at fixed power without considering their channel gain, which results in strong interference to GB users as well as increased interference among them, which negatively affects the throughput of both types of users. Allowing more than 12 GF users access drastically reduces GB users' throughput while also lowering GF users' throughput due to high interference. Further increasing the number of users may result in decoding failure of GB users. In contrast, both types of users achieve significant throughput and can accommodate more users while satisfying the QoS of GB users.

\subsection{Scalability of the Proposed Algorithm}
One of the major issues with MA-DRL algorithms is the scalability. One solution to this problem is to use decentralised learning with networked agents, which enables agents to communicate information about their actions, states, or policies with other agents \cite{li2021applications}. However, such communication among agents increases communication overhead and reduces energy efficiency. To reduce this communication overhead, in our proposed algorithm, the agents indirectly receives information (data rate) of other users form the BS as state and updates their policies in a decentralized manner. More specifically, agents (users) are independent learners, and they cannot communicate with each other directly, which is preferred for applications involving high communication costs or, unreliable communications for example, in UAV networks or IoT networks. We depicted the scalability of our proposed algorithm in Fig. \ref{7}(\subref{sc}), where one can see that our proposed algorithm converges almost in the same number of episodes against different number of users. With $\mathbf{N}=9$ (3 GF users per sub-channel) agents, the algorithm starts converging in about 255 episodes and converges to its maximum reward value in about 325 episodes. A similar performance can be seen when we increased the density, i.e., number of agents to $\mathbf{N}=12$ (4 GF users in each sub-channel), $\mathbf{N}=15$ (5 GF users in each sub-channel), $\mathbf{N}=24$ (8 GF users in each sub-channel) and $\mathbf{N}=30$ (10 GF users in each sub-channel). Increasing this number further makes it difficult for power-domain NOMA to successfully decode more than 10 users in a NOMA cluster. Furthermore, new agents can be added to the existing trained agents by simply copying the NN parameters of the trained agents, which can generalize the proposed method to diverse scenarios. Therefore, our proposed algorithm is suitable for SGF-NOMA IoT networks with a massive number of users.
\begin{figure}
	\centering
	\begin{subfigure}[b]{0.5\textwidth}
		\includegraphics[width=1\linewidth]{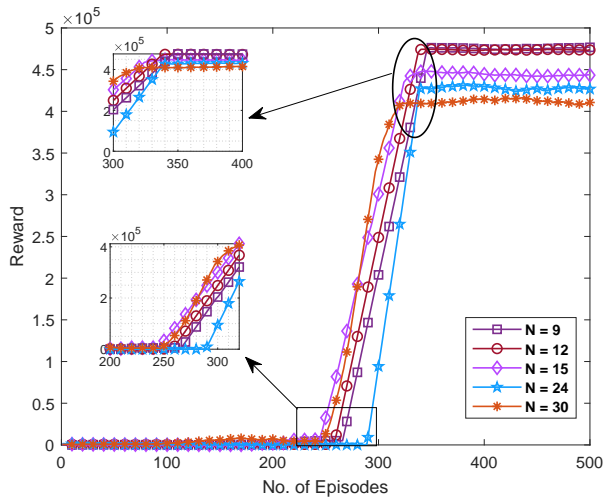}
		\caption{Convergence w.r.t different number of agents}
		\label{sc}
	\end{subfigure}
	
	\begin{subfigure}[b]{0.5\textwidth}
		\includegraphics[width=1\linewidth]{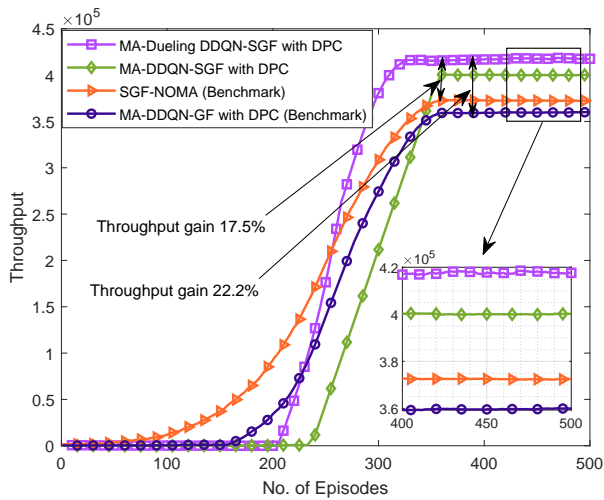}
		\caption{Performance comparison}
		\label{scb}
	\end{subfigure}
	\caption[7ab]{The scalability and performance comparison: Sub-figure (a) shows the scalability of our proposed algorithm with increasing number of agents. Sub-figure (b) shows the performance comparison of the proposed MA-Dueling DDQN based SGF-NOMA and MA-DDQN based SGF-NOMA with pure GF-NOMA scheme\cite{fayaz2021transmit} and SGF-NOMA \cite{8662677}.}
	\label{7}
\end{figure}
\subsection{System Performance Comparison}
Fig. \ref{7}(\subref{scb}) shows the performance gain of the proposed MA-DDQN based SGF-NOMA and MA-dueling DDQN based SGF-NOMA with DPC mechanism compared to other methods, namely pure GF-NOMA and SGF-NOMA\cite{8662677}. It can be seen that MA-DDQN based SGF-NOMA and MA-dueling DDQN based SGF-NOMA with DPC mechanism outperform the benchmarks in terms of throughput. The proposed MA-dueling DDQN based SGF-NOMA achieves 22.2\% and 17.5\% more throughput than the pure GF-NOMA and SGF-NOMA, respectively. This is because in our proposed algorithm, only a subset of GF users are allowed to transmit on a sub-channel exclusively occupied by GB users, which causes interference within a tolerable threshold to GB users.
Unlike the benchmark scheme mentioned in \cite{8662677}, where all GF users transmit at a fixed power regardless of their channel gain, in our proposed algorithm, GF users distributively choose the transmission power according to their channel gain and geographical location. Since each GF user acts as an agent trying to maximize the reward (network throughput). Therefore, GF users choose the transmit power level that causes less interference (or control interference) to other users and form a NOMA cluster which increases the system throughput. Similarly, in the pure GF IoT network, all GF users are allowed to transmit data, which leads to strong interference on the sub-channels, increasing the intra-RB interference and resulting in low network throughput. 
\vspace{-0.8cm}
\subsection{PP Associated to each Sub-Channel and Comparison with Conventional Open-Loop PC}
In section-V (D), we concluded that at most four users on each sub-channel (cluster) provide the highest cluster and system throughput. Thus, we design a PP for each sub-channel with an upper limit that no more than four GF users can access a sub-channel occupied by GB users. The designed PPs associated with each sub-channel is depicted in Fig. \ref{op}(\subref{pps}). We used three sub-channels that are pre-occupied by GB users and available to GF users for uplink transmission. From the figure, it can be seen that we have found the optimal transmit power levels for each PP and mapped them to the corresponding sub-channels. The BS broadcasts PPs and other information to all GF users in the network. After receiving this information, GF users select the transmit power randomly from the PP associated with the selected sub-channel. Therefore, the proposed SGF-NOMA provides an open-loop PC with less signalling overhead and small latency. Next, we use the fixed power allocation (FPA) PC method is an open-loop PC algorithm as baseline. Form Fig. \ref{op}(\subref{opn}), it can be seen that our proposed open-loop PC mechanism outperforms the conventional open-loop PC approach. Because all the users in the conventional approach uses the same power level for uplink transmission. On the other hand, our proposed approach provides multiple power levels that are associated to different sub-channels and users can select the best power level according to their channel gains and geographic locations while satisfying the QoS requirements of GB users.
\begin{figure}[h]
	\centering
	\begin{subfigure}[b]{0.5\textwidth}
		\includegraphics[width=1\linewidth]{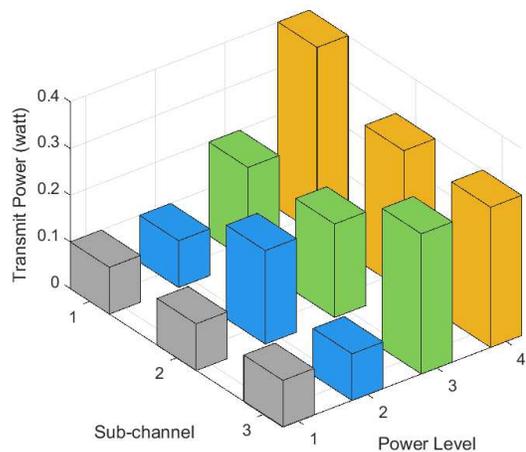}
		\caption{PP associated to each sub-channel}
		\label{pps}
	\end{subfigure}
	\begin{subfigure}[b]{0.5\textwidth}
		\includegraphics[width=1\linewidth]{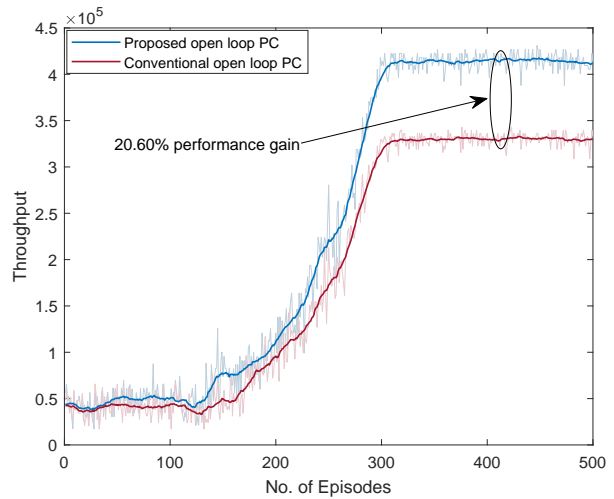}
		\caption{Conventional open-loop PC vs. proposed}
		\label{opn}
	\end{subfigure}
	\caption[8ab]{The designed PPs and comparison with conventional open-loop PC. Sub-figure (a) shows the PP associated to corresponding sub-channel. Sub-figure (b) illustrates the performance comparison in terms of throughput with conventional open-loop PC method.}
	\label{op}
\end{figure} 
\section{Conclusions and Future Research Directions}
In this paper, we have proposed MA-DRL based SGF-NOMA algorithm with DPC to generate PP and map it to each sub-channel, which enables open loop power control. In the proposed scheme, a single user is granted access to the sub-channel through a GB protocol and GF users are admitted to the same sub-channel via GF protocol. We have verified that our proposed algorithm is scalable for SGF-NOMA IoT networks with a large number of users. Numerical results show that the proposed MA-DRL-based SGF-NOMA with DPC outperforms the SGF-NOMA system and networks with pure GF protocols with 17.5\% and 22.2\% gain in system throughput, respectively. We have shown that the proposed algorithm is computationally scalable regardless of the number of users and efficiently allocates the additional data rate of GB users to GF users, which improves connectivity. Additionally, to enhance system throughput, we also determined the upper bound of GF users that can access a sub-channel reserved by GB users. Investigating the environmental impact on the considered DRL algorithms and user fairness (in terms of energy consumption) are the promising future directions. 

\singlespacing
\bibliographystyle{IEEEtran}
\bibliography{mybib}

\end{document}